# Automated Single-Label Patent Classification using Ensemble Classifiers

ELENI KAMATERI

VASILEIOS STAMATIS

KONSTANTINOS DIAMANTARAS

MICHAIL SALAMPASIS

Department of Information and Electronic Engineering, International Hellenic University, Thessaloniki 57400, Greece

Many thousands of patent applications arrive at patent offices around the world every day. One important subtask when a patent application is submitted is to assign one or more classification codes from the complex and hierarchical patent classification schemes that will enable routing of the patent application to a patent examiner who is knowledgeable about the specific technical field. This task is typically undertaken by patent professionals, however due to the large number of applications and the potential complexity of an invention, they are usually overwhelmed. Therefore, there is a need for this code assignment manual task to be supported or even fully automated by classification systems that will classify patent applications, hopefully with an accuracy close to patent professionals. Like in many other text analysis problems, in the last years, this intellectually demanding task has been studied using word embeddings and deep learning techniques. In this paper we shortly review these research efforts and experiment with similar deep learning techniques using different feature representations on automatic patent classification in the level of sub-classes. On top of that, we present an innovative method of ensemble classifiers trained with different parts of the patent document. To the best of our knowledge, this is the first time that an ensemble method was proposed for the patent classification problem. Our first results are quite promising showing that an ensemble architecture of classifiers significantly outperforms current state-of-the-art techniques using the same classifiers as standalone solutions.

## 1 INTRODUCTION

Patents are an important economic asset and patent filing rates have enormously increased. Patent offices internationally must deal with these large number of applications. Therefore, automating any subtask of the large patent examination process is an important challenge which has significant impact because it can speed up the examination process. One subtask that can be automated is the pre-classification which is typically done by front-line patent experts who manually classify an application using a classification scheme (e.g., IPC, CPC). This task is quite important considering that this classification will route the patent application to a sub-department of the office for detailed examination. Also, during the examination process, properly assigning other relevant classification codes ensures that patents with similar technical features will be grouped under the same intellectual scheme, something which is crucially important for many subsequent patent retrieval tasks. For example, in prior art search, the assigned codes could substantially improve the search for relevant patents in different ways such as filtering the search or expanding with extra search terms.

Classification schemes follow a hierarchical structure meaning that each inner node in the hierarchy has exactly one parent and the path to each code is unique. The most widely used classification scheme is the International Patent Classification (IPC) which follows this hierarchical structure containing thousands of codes each representing a more general (at higher levels) or very specific technological concepts. In 2013, the Cooperative Patent Classification (CPC) has been introduced by EPO and USPTO as a common classification scheme that contains approximately 250,000 individual codes organized in a similar hierarchical structure. In the version of IPC 2006 for example, the classification scheme contains in total 8 sections, 131 classes, 642 sub-classes, 7,537 groups and 69,487 subgroups. Furthermore, the classification schemes are periodically updated meaning that there is a substantial need to support the potential re-classification of some patents.

Two important features for automated classification are the following: a) it can be done at different levels of the classification hierarchy and b) it can be done either as a single-label task in case of the pre-classification stage (described above) or as a multi-label task because each patent can have multiple codes assigned, sometimes at a different level of the hierarchy. Several research works tried to automate the patent classification task, however because automated systems do not attain human performance, the automation only worked at a higher level of the hierarchy.

Recently, automated patent classification has been examined using deep learning (DL) methods such as convolutional neural networks (CNN) [6, 13, 14], Word2Vec and Long-short term memory (LSTM) [4, 5, 10] and other DL methods [7] to predict the most representative classification code(s) for a patent. In the current work, we evaluate these CNN and RNN models employing different language models and different parts of the patent document. Additionally, we propose an ensemble method combining the performance of standalone CNN and RNN classifiers trained at different parts of the patent document. This study focuses on single-label classification at the sub-class (3rd) level category of the IPC 5+ level hierarchy. We selected the sub-class level because this is the level in which the pre-classification task is typically performed. Moreover, the sub-class level is the minimum level in the IPC hierarchy where it has value to use an automated system, although the group (4th) and sub-group (5th) level are even more useful. However, it should be noted that classifying patents at the sub-class is already a very difficult task, because at this level there are approximately 650 labels, much more than those found in a typical text classification problem. Furthermore, the classification task becomes even more complex considering that only one main label should be accurately assigned to each patent.

To create a proper dataset for evaluating our methods for single-label classification, we use the <main classification> tag that exists in patents in the CLEF-IP 2011 dataset to extract the primary classification. This label denotes the main and most important classification code assigned to a patent. Keeping only the patents that have a main classification tag assigned, a subset of the CLEF-IP collection is created.

The remainder of the paper is structured as follows: Section 2 describes related work in the area of automatic patent classification. Section 3 describes the proposed ensemble method and the individual CNN and RNN classifiers taking place in the ensemble architecture. Section 4 presents the experimental methodology and setup to evaluate our method. Section 5 presents and discusses the experimental results. Finally, section 6 concludes the paper and discuss the ongoing and future work.

## 2 PRIOR WORK

Several works were presented to solve the automated patent classification problem. Actually, patent classification methods advance hand-to-hand with the text and document classification that are typical Natural Language Processing (NLP) applications which deal with the assignment of one or multiple pre-defined labels to a given text or document respectively [14].

Earlier research on patent classification applied basic NLP and feature engineering techniques to pre-process texts before feeding them into well-known classifiers. For example, Fall et al. [16] performed stop words removal, stemming, term selection using the information gain and then fed the transformed texts into Naïve Bayes (NB), Support Vector Machine (SVM), and K-Nearest Neighbor (KNN) classifiers. Tikk et all. [2] applied stop word removal, stemming, dimensionality reduction and removal of rare terms, and they sent the processed output into a neural network, named HITEC, which copies the tree-like structure of the taxonomy. Their method showed 53.25% classification accuracy at the sub-class level. Similarly, Lim and Kwon [11] constructed a stop words list dedicated to patent domain while they selected their feature set using a TF-ICF weighing scheme, which is a variation of the well-known TF-IDF.

Around 2017, research on automated patent classification turned to test out the effectiveness of new DL methods for text processing tasks. Among them, Grawe et al. [4] used stop word removal and then transformed the cleaned text into a meaningful representation produced by the Word2Vec that was sent as input to an LSTM network (a type of RNN). The method achieved 63% accuracy at the sub-class level. Likewise, Xiao et al. [5] used Word2Vec embeddings and LSTM to classify patents in the security field.

Another trend that is recent in the relevant literature is the training of Word Embeddings on domain-specific patent datasets. Risch and Krestel [7] used the FastText embedding that was trained on a patent dataset together with bi-directional GRUs (another type of RNN) to achieve better performance compared with Word Embeddings trained on Wiki documents. Their method achieved 53% accuracy at the sub-class level. Moreover, Sofean [10] developed a self-trained Word Embedding trained on million patents and then used a LSTM network to perform patent classification. For evaluating his method, Sofean kept only the sub-classes appearing in more than 500 documents, resulting to 43 sub-classes, and obtained an accuracy of 67%.

Another improvement in patent classification with respect to the DL techniques started with the adoption of CNNs. Li et al. [6] proposed a DL algorithm, DeepPatent, for patent classification by combining word vector representation and a well-designed CNN. DeepPatent performs multi-label classification at the sub-class level getting 75.46% recall at top 4. Likewise, Zhu et al. [13] used the Word Embedding technique to segment and vectorize the input data and then a symmetric hierarchical CNN, named PAC-HCNN, to classify patents outperforming traditional RNN. Moreover, Abdelgawad et al. [14] compared several recent neural network models and showed that CNNs are a suitable choice for patent classification. The authors also proved that state-of-the-art hyperparameter optimization techniques can further improve the CNN performance getting an accuracy of 52.02 at the sub-class level.

Recent trends on DL that includes pre-trained unsupervised language models on large corpuses and fine-tuning them on downstream tasks have produced state-of-the-art performances. Among them, the BERT model, released by Google in 2018, is the first deeply bidirectional, unsupervised language representation, pre-trained using only a plain text corpus in order to infuse later the context in which it will be used. Following this trend, Lee et al. [15] leveraged and fine-tuned the BERT-Base model and applied it to patent classification getting better accuracy than other recent DL approaches. This method was evaluated on a multi-label classification and achieved 54.33 recall at 1 Similarly, Roudsari et al. [8] presented a short work applying and fine-tuning Distil-BERT model for patent classification.

Innovative architectures were also proposed working on feature extraction from patent text. Bai et al. [9] proposed a multi-stage feature extraction network, named MEXN, which divides the document to fixed-sized paragraphs and extracts features, then summarizes the paragraph features into a document feature using the attention mechanism, and finally computes the weight matrix with hierarchical category information of the patent. Similarly, Hu et al. [12] proposed a hierarchical feature extraction model, named HFEM, which is able to capture both local features of phrases as well as global and temporal semantics.

Another important aspect that is worth mentioning is the limited number of works considering the hierarchy of the IPC/CPC codes. In principle, flat methods implement classification using only a single level (e.g. sub-class), while hierarchical methods exploit the knowledge of hierarchy (previous associated codes at different levels) to predict an effective classification. Although flat methods seem to dominate the literature in patent classification techniques, there are still some outstanding examples. Chen and Chang [1] presented a novel patent classification method that combines flat text classification in two different levels of the IPC hierarchy with clustering method to perform patent classification down to the subgroup level. In [17], Giachanou et al. exploited the hierarchical structure of IPC codes to propose a multi-layer collection selection method that can be seen as a patent classification problem. Their method, in addition to utilizing the topical relevance of IPCs at a particular level of interest, exploits the topical relevance of their ancestors in the IPC hierarchy and aggregates those multiple estimations of relevance to a single estimation. Furthermore, Risch et al. [3] used a sequence-to-sequence neural network with label embeddings to generate a sequence of labels for all levels where the embeddings of previous level are considered for generating the next labels. This method achieved 56.7 accuracy at the sub-class level.

Last but not least, some researchers tried to identify which parts in a patent document can provide more representative information for classification tasks [11, 15]. We distinguish the use of technical and background parts extracted from the descriptions section [11], the first claim [15], the title and abstract [6, 13], or at most cases, the title, abstract claims and description [7, 10, 12, 14].

## 3 METHODS

An ensemble method receives evidences from multiple models, working either at the same or different sources of information, combines these evidences and produces a final prediction. This technique obtains better predictive performance than that could be obtained from any of the constituent models [18] since it exploits potentially not related information coming from all single models. Although ensemble methods are experiencing good results in many applications, they are less explored for automated patent classification and only for upper levels of the IPC hierarchy [19].

In this section, we describe the ensemble architecture we propose and the individual neural network classifiers that we deployed to address the automatic patent classification problem down to the sub-class level. Although the individual CNN and RNN classifiers can further be improved by tailoring their structure with additional layers and fine-tuning the different parameters, we selected to present here the results achieved from simple versions of the selected neural networks because the accuracy rates are not significantly improved when more complex architectures were tested.

## 3.1 Ensemble architecture

The ensemble architecture consists of three individual classifiers with each of them trained on a different part of the patent text, i.e., the title-abstract, the description and the claims section, respectively. Each classifier produces a list of probabilities for all labels based on its partial knowledge about the patent. Then, the probabilities for a specific label derived from the three individual classifiers are averaged and a final probability is calculated for this label. The label with the maximum probability consists the predicted/expected label for the patent. Figure 1 depicts the proposed ensemble architecture.

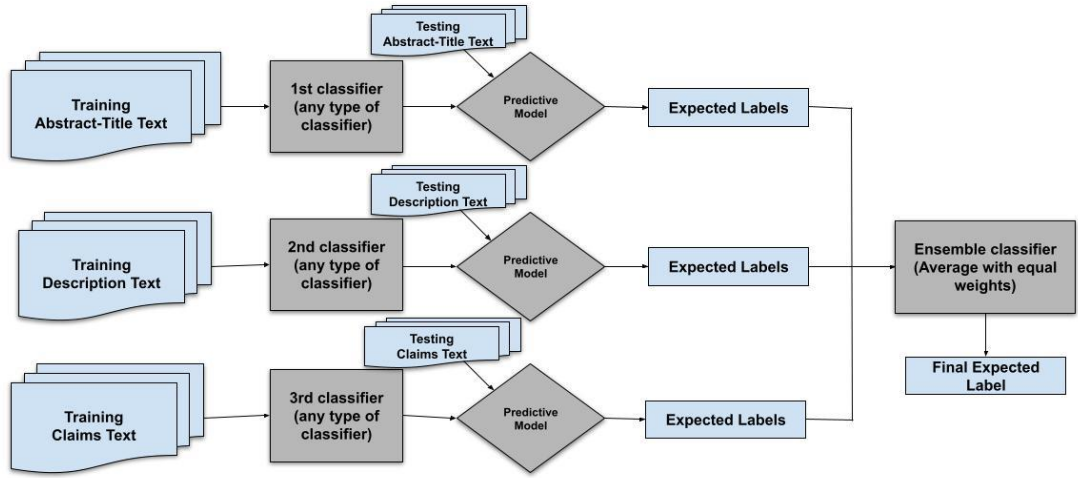

Figure 1: An ensemble architecture of classifiers

## 3.2 Individual classifiers

As mentioned above, the ensemble architecture combines the outcome of individual classifiers. In our study, state-of-the-art CNN and RNN classifiers are used and evaluated since, based on the current literature review, they are experiencing good results in automated patent classification.

### 3.2.1 CNN

CNN is one of the neural networks that have been widely used recently in the domain of automated patent classification. In our method, the pre-processed text is fed to a CNN through an embedding lookup, which converts words to vectors represented in a high-dimensional vector space. Afterwards, a 1-D convolutional layer is applied on top of the embedding layer. A max-pooling along with a flatten layer are then applied sequentially to the output of the convolutional layer. After a dense of 1024 filters and a dropout layer with a dropout rate of 0.5, the output is fed in another dense layer with a softmax activation in order to obtain a probability distribution over all targeted labels (IPC sub-classes which are equal to 659).

### 3.2.2 RNN

RNN is another category of neural networks that have been widely used in text classification. Both LSTM and GRU have been proposed as variations of RNN to deal with the exploding gradient and vanishing gradient problems of back propagation through time experienced by RNN models. Since they consistently produce good results in text analysis tasks, they have been applied in many fields including automated patent classification. Here, we evaluate both LSTM and GRU models as they are superior to simple recurrent units. Moreover, we use bi-directional LSTM and GRU so that the input sequence to be processed in two directions of the sentence. Similar with CNN architecture, we convert each word to its embedding. This embedding layer of words is processed by a spatial dropout, which randomly masks 10% of the input words to make the neural network more robust. The remaining 90% serves as input to the LSTM or GRU layer (also to the bidirectional LSTM and bidirectional GRU). The output of these steps is inserted in a dense layer with as many units as the targeted labels and a softmax activation.

## 4 EXPERIMENTS

Here follows the description of the evaluation methodology for assessing our methods. More specifically, we describe the data collection, the selected data pre-processing, the experimental methodology, the different sets of experiments that we conducted to address the automatic patent classification problem in the sub-class level, and the metrics for evaluating our experiments.

### 4.1 Data collection

For evaluating our techniques, we used the CLEF-IP 2011 test collection. From all patent documents in the test collection, we kept the documents that have English text, pertaining to 1,149,536 patent documents. From those patents, only 418,788 have the required information of which is their main classification category, i.e. include the <main classification> tag. This information was mandatory to determine which is the primary classification code, so a single-label classification method can be reliably executed and evaluated.

From the initial 418,788 only 302,578 patent documents have been remained as only these contain a title-abstract, a description and a claims section in the patent document at the same time.

From those patents, we created four different pools of patents: i) the initial pool which contains all text from the aforementioned sections; ii) a second pool which contains the title and the abstract section; iii) a third pool which contains the description section and; iv) a last pool which contains the claims section. These different pools have been created for testing whether using different sections will have an effect on the classification task.

Table 1 shows some basic statistics regarding the number of words per each section in the sub-dataset of 302,578 patent documents that we used:

Table 1: Statistics of different patent parts

|               | Title-Abstract | Description | Claims |
|---------------|----------------|-------------|--------|
| Min num words | 2              | 1           | 1      |
| Max num words | 7,692          | 12,030      | 61,972 |
| Mean num words| 102            | 457         | 1,044  |

### 4.2 Data pre-processing

Afterwards, we extracted the first "X" words from the concatenated result of the title-abstract, the description, and the claims sections. This means that most of the words used to represent the patent document are taken from the title-abstract section since the average number of words from the title-abstract section is 102 and can reach 7,692 words at the maximum (Table 1). This leaves out a (often small) number of words which comes from the description section, while words from the claims section will be not used in the final "X" words length representation of the patent document. We repeated the same methodology in rest three "mono-thematic" data collections retrieving the first "X" words coming from title and abstract, the description and the claims sections for each patent, respectively. For all data collections, the different number of words, called "X", that has been tested was: {20, 40, 60, 80, 100, 200, 300, 400}.

The method of retrieving the first "X" words for the representation of a patent document is followed in several literature works [4-7]. Although we followed this method for the current experimental setup, we have introduced some improvements, such as the usage of first "X" words of a patent part. Another improvement is the usage of first "Y" words from all patent parts.

### 4.3 Experimental methodology

For each patent, the patent target variable (i.e., sub-class IPC code) was encoded using one-hot encoding. On the other hand, the patent text, formed by the selected first "X" or "Y" words, was followed a sequence of processing steps. It was tokenized and converted into a sequence of tokens, while padding was used to ensure equal length vectors. Tokens were then mapped to embeddings and an embedding matrix representing each patent document has been created. For the transformation of tokens to embeddings, we used the pre-trained language models of FastText, Word2Vec, and Glove, with dimension size set to 300 for the generated word vector. We also trained Word2Vec using the genism toolkit in our corpus (the corpus of 302,712 patent documents) and then we used the produced embeddings to represent the patent texts. For the training of Word2Vec, we set the vector size to 300, the maximum distance between current and predicted word to 8 and the number of iterations to 20. The outcome embedding matrix was later used as input to the neural network architecture, which consists of several layers and returns as outcome a probability for each target.

Moreover, the dataset of 302,712 patents was split into training, validation and testing sets; 80% for training, 10% for validation, while other 10% was kept out for testing the classification model.

After some tests, we set batch size to 128 and epochs to 5 for CNN models and 15 for RNN models.

### 4.4 Experimental setups

This experimental layout explores the effect of different deep learning models, language models and feature selection methods on patent classification results. More specifically, we evaluated how different batches of words coming from different parts of the patent document may result into different outcomes on patent classification (Exp#1). Then, we tested how different embedding representations can affect the patent classification outcomes (Exp#2). Moreover, we also explored the effect of different deep learning models on patent classification results (Exp#3). Last, we evaluated the ensemble method of combining the results of similar classifiers trained in parallel in the three different parts of the patent document (Exp#4).

### 4.5 Evaluation criteria for single-label patent classification

For each experiment, we used the following evaluation metrics to evaluate our methods. The selection of the specific metrics was made after a thorough review of the literature which showed that these are the only metrics that accurately resembles the actual real task of single-label patent classification. Firstly, we predicted the sub-class IPC codes for each patent document. Then we calculated *Accuracy* and *Recall at n* for each experiment.

$$Accuracy = \frac{\# \ of \ correct \ predictions}{\# \ of \ patents \ in \ the \ testing \ set}$$

With *Accuracy* metric, we evaluate the first prediction of the classifier for each patent document. If it is a correct relevant result, the numerator is increased by one, while the denominator is equal to the number of patent documents in the testing set.

$$Recall\ at\ n = \frac{correct\ predictions\ up\ to\ position\ n}{\#\ of\ patents\ in\ the\ testing\ set}$$

For *Recall at n* metric, we evaluate the top-n predictions for each patent document. If one of them is the correct relevant result, the numerator is increased by one. The denominator is equal to the number of patents in the testing set.

## 5 RESULTS

### 5.1 Exploring different feature selection techniques for patent representation

In this set of experiments (Exp#1), we extracted the first "X" words of a patent part (retrieving the patent text stored in each data collection) and the first "Y" words from all patent parts (retrieving equal words coming from the title-abstract, the description and the claims part) for the representation of the patent document. Moreover, we used the CNN architecture for the classification task and the FastText pre-trained word embedding for the text representation. Figure 2 illustrates the accuracy scores for varying numbers of retrieved words from a specific patent part each time, compared to the accuracy achieved when we retrieve words from each patent part simultaneously.

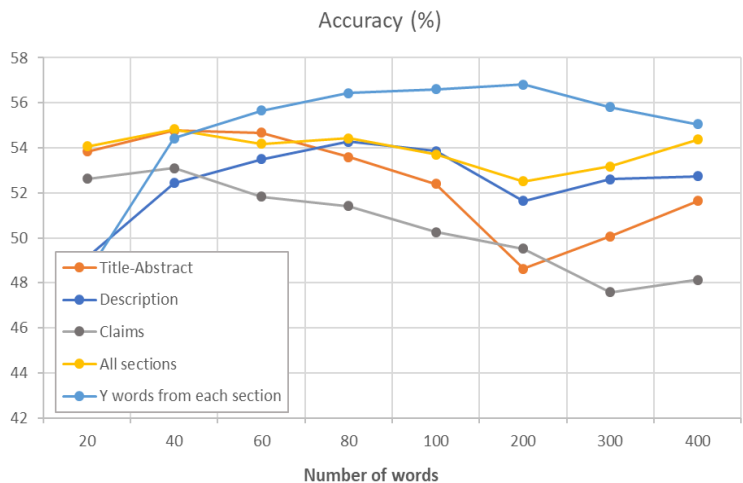

Figure 2: Graphical representation of accuracy scores achieved for a varying number of words retrieved each time from a different data collection (first four lines) in comparison with the accuracy score achieved for a varying number of words retrieved in parallel from all patent parts (light blue line labelled "Y words from each section").

The accuracy is better when retrieving "Y words from each section" at the same time to represent the patent document. After, it comes the accuracy when retrieving words from "all sections", then follows the accuracy of the "title-abstract", the "description", and last comes "the claims". Therefore, the best representation is achieved when using words from all sections. Moreover, the title-abstract and description sections are constantly important descriptors of the patent document. The claims sections have the lowest accuracy score meaning that these have low representation value in the patent document which is reasonable considering that this section defines the legal boundaries of an invention using difficult and vague vocabulary.

The scores for "all sections" and "title-abstract" section are initially quite similar, which is reasonable considering that the number of words is low and most or even all of the words that are retrieved from the data collection containing text from all (concatenated) sections are totally coming from the title-abstract section. For bigger number of words, we observe that the accuracy scores for "all sections" and "description" section become almost equal. This can be interpreted that the first words used in the description section are quite important for the patent representation.

## 5.2 Exploring different language models

In the current experimental setup (Exp#2), we explored how different word embeddings can affect the patent classification outcome. For the experiment, we retrieved the first 60 words for representing each patent document and used the CNN architecture for classifying them. In Figure 3, the accuracy scores for different language models and data collections are displayed.

Among pre-trained language models, the FastText seems to achieve better representation of the patent text and thus better accuracy scores than other two pre-trained language models (i.e., Word2Vec and Glove). For the FastText, the accuracy reaches 52.69% for the dataset of all sections. It is also interesting that even when the Word2Vec language model has been trained on the domain-specific corpus and then used for the representation of the patent text, it reaches worse results (52.01%) than FastText.

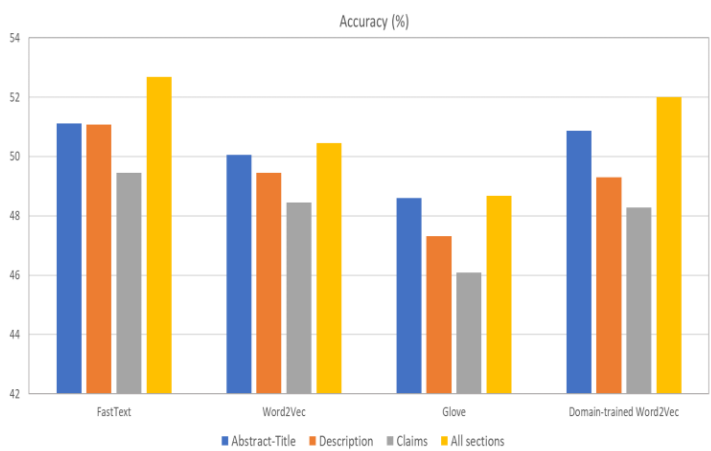

Figure 3: Graphical representation of accuracy scores using different language models for representing the patent document. The embeddings has been created considering the first 60 words coming either from i. the title-abstract section; ii. the description section, iii. the claims sections and iv. the concatenation of all the above sections.

## 5.3 Exploring different DL methods

In this subsection, we explored the effect of different deep learning models on patent classification results (Exp#3). As mentioned above, the DL algorithms that were selected to be evaluated are a CNN, a Bidirectional LSTM, a Bidirectional GRU, a LSTM and a GRU. We also investigated whether a simple ensemble method of averaging the outcome of three independent classifiers applied on the three "mono-thematic" data collections can achieve better classification results than each of those classifiers acting on its own. For our experiment, the first 60 words were retrieved for all data collections and the FastText pre-trained word embedding was used for the text representation. In Figure 4, we can see the accuracy scores achieved for different DL models for all data collections.

It is shown that LSTM and GRU models perform similar, while a significant improvement in accuracy is achieved by the usage of the bidirectional DL models (either GRU and LSTM), which demonstrate the best accuracy scores among all data collections. More specifically, the bidirectional LSTM reaches the best accuracy score of 59.83% for the data collection of the title and abstract.

The bi-directional LSTM achieves better classification performance than almost all similar state-of-the-art methods [15, 14, 7, 3], except for [4] which exploits a language model trained on a domain-specific dataset.

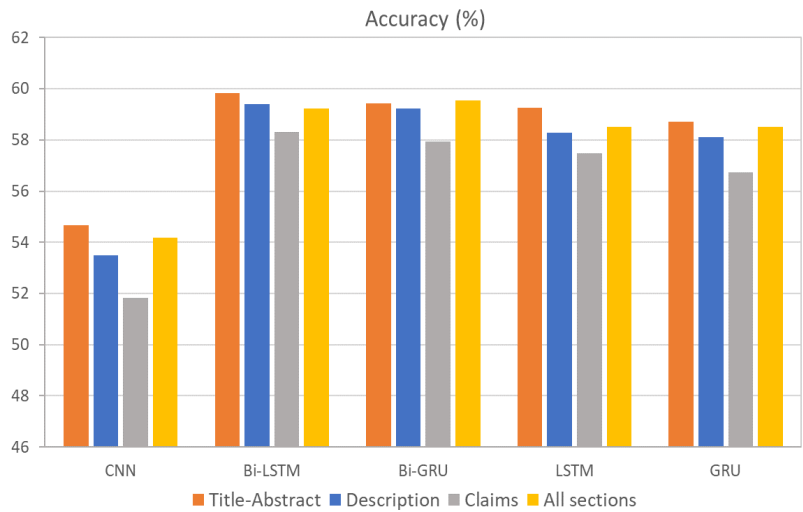

Figure 4: Graphical representation of accuracy scores using different DL models for patent classification. The DL models have been evaluated for the first 60 words coming either from i. the title-abstract section; ii. the description section, iii. the claims section or iv. the concatenation of all the above sections.

### 5.4 Ensemble method

In the current experimental setup (Exp#4), we investigated whether a simple ensemble method of averaging the outcome of three identical classifiers applied on the three "mono-thematic" data collections can achieve better classification results than each of those classifiers acting on its own.

Figure 5a presents the evaluation metrics of each individual classifier trained on the 60 words coming from the title-abstract part, the description part and the claims part, respectively, and (with blue) the evaluation scores of the ensemble method combining the results of the above three individual classifiers working on separate parts of the patent text.

We observe that all evaluation scores are much improved when we apply the ensemble method compared with individual classifiers. More specifically, the ensemble of bidirectional GRU classifiers performed on different patent parts achieved an accuracy of 65%, which is the best accuracy score compared with the rest combinations of classifiers.

Figures 5b, 5c, 5d show the recall at first three, at first five and at first ten returned results, respectively, for individual classifiers and the combination (ensemble) of them.

The most interesting observation in these figures is that the recall at n (recall) increases significantly when we are looking at the top three, five and ten returned targets. Specifically, the recall at the top three returned targets reaches 85.88% (optimal R@3) when we used bi-directional LSTM as individual clarifiers and training them on different patent parts. The ensemble classification model will predict the correct label among the top five returned with an accuracy of 91.35% (optimal R@5) and will predict the correct label among the top ten returned with an accuracy of 95.65% (optimal R@10).

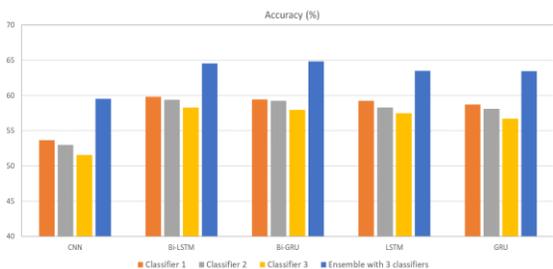
Figure 5a

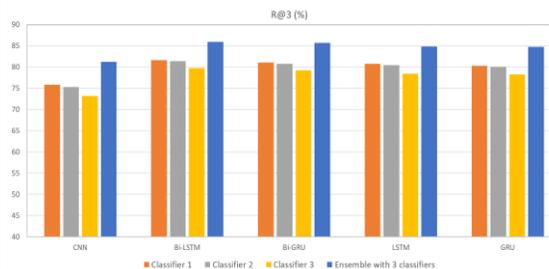
Figure 5b

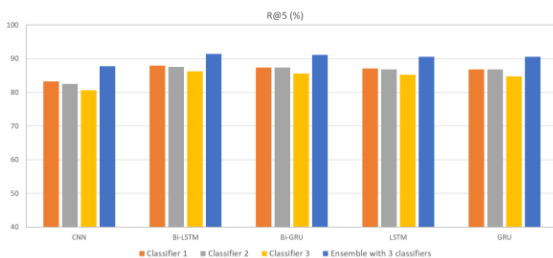
Figure 5c

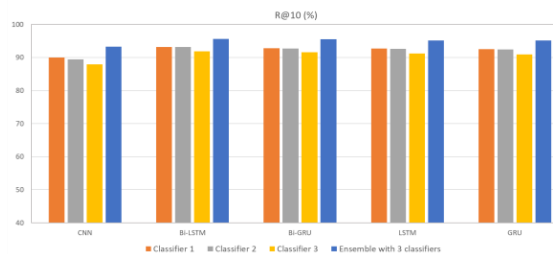
Figure 5d

Figure 5a, b, c, d: Accuracy, Recall@3 (R@3), Recall@5 (R@5) and Recall@10 (R@10) scores using individual classifiers and an ensemble method of combining these individual classifiers.

In order to quantify the improvement experienced with ensemble method, we present in Table 2 the average accuracy achieved by all individual classifiers compared with the accuracy achieved by the ensemble method. The highest increase was observed when we applied the ensemble method with CNN individual classifiers (13%), followed by the ensemble method with bidirectional GRU individual classifiers (10%).

Table 2: Improvement in accuracy scores achieved by ensemble method.

|  | CNN | Bi-LSTM | Bi-GRU | LSTM | GRU |
|---|---|---|---|---|---|
| Accuracy of **individual classifier 1** applied on title-abstract | 53.65 | 59.83 | 59.43 | 59.26 | 58.71 |
| Accuracy of **individual classifier 2** applied on description | 52.99 | 59.40 | 59.24 | 58.28 | 58.10 |
| Accuracy of **individual classifier 3** applied on claims | 51.54 | 58.31 | 57.93 | 57.48 | 56.72 |
| **Average accuracy** of individual classifiers applied 1, 2 and 3 | 52.73 | 59.18 | 58.87 | 58.34 | 57.84 |
| **Ensemble method** combining individual classifiers 1, 2 and 3 | **59.54** | **64.57** | **64.85** | **63.51** | **63.44** |
| **Improvement (%)** of ensemble method compared to averaged accuracy of individual classifiers | 12.92 | 9.11 | 10.16 | 8.87 | 9.69 |

Moreover, the ensemble of three identical classifiers trained with different patent text achieves better classification accuracy compared to the results provided by state-of-the-art research efforts [3, 4, 7, 14, 15].

## 6 CONCLUSION

Our work on automated single-label patent classification was mainly driven by our need to experiment and become familiar on how DL models could be used and combined to develop tools that can effectively support the pre-classification task in patent offices. Additionally, we wished to better understand how various parameters of the DL models and domain specific parameters, such as the selection of patent document sections/feature words and language modelling representations, may influence automated patent classification.

The evaluation of different DL models showed that a bidirectional-LSTM or a bidirectional-GRU can achieve better results than other DL methods, especially when it is combined with FastText word embeddings. Moreover, an ensemble method can further increase the classification performance. The highest accuracy score of 65% achieved when an ensemble architecture of three bidirectional GRU classifiers was used with each of them getting as input the title-abstract, the descriptions and the claims section, respectively.

Patent sections from which the feature words are retrieved and words selected to represent the patent document also seem to play an important role for automated patent classification. Although the abstract section offers the most valuable words for automated patent classification, the description section contains also important information that makes a patent distinguishable. Moreover, the selection of feature words from all parts of a patent at the same time achieves better results than the selection of the first words from each part.

We believe that the initial work we presented in this paper clearly illustrates that an ensemble method using the optimal DL models and domain parameters would be useful to support patent professionals in a pre-classification task. For example, our ensemble method produces a promising result when the top 3 or 5 recommended classification codes are predicted.

Our plan is to continue this work on the ensemble method of different classifiers by focusing on better optimizations of DL models, better representations of patent documents and better exploration of language models. We would like also to further experiment with different strategies for combining the results of different classifiers in our ensemble architecture and to further analyze the rankings that are produced from various classifiers. Also, we would like to utilize the hierarchical structure of IPC codes and the description provided for each of them. Last, we plan to extend these approaches on multi-label patent classification and use the complete CLEF-IP test collection that might also resolve the over-fitting problems of DL models.

In conclusion, we feel that in this paper we have already produced some initial knowledge and useful results which will help engineers to produce effective patent classification tools for patent pre-classification or to support other patent retrieval tasks.

## ACKNOWLEDGMENTS

This work has received funding from the EU Horizon 2020 research and innovation programme under the Marie Skłodowska-Curie grant agreement No: 860721 (DoSSIER Project, https://dossier-project.eu/).